\documentclass[prl,twocolumn,groupedaddress,nobalancelastpage,floatfix]{revtex4}
\usepackage{graphicx}

\begin{document}

\title{Role of c-axis pairs in V$_{2}$O$_{3}$ from the band-structure point
of view.}
\author{I.S.~Elfimov$^1$, T. Saha-Dasgupta$^{2}$ and M.A.~Korotin$^{3}$}
\affiliation{$^1$Department of Physics and Astronomy, University of British
Columbia, 6224 Agricultural Road, Vancouver, B.C. V6T 1Z1, Canada\\
$^2$S.N.~Bose. National Center for Basic Sciences JD~Block, Salt Lake,
Kolkata 700 098, India\\
$^3$Institute of Metal Physics, 620219 Ekaterinburg GSP-170, Russia\\
}

\begin{abstract}
The common interpretation of the LDA band structure of V$_{2}$O$_{3}$ is
that the apparent splitting of the $a_{1g}$ band into a low intensity
structure deep below the Fermi energy and a high intensity feature above it,
is due to the bonding-antibonding coupling of the vertical V-V pair. Using
tight-binding fitting to --as well as first-principles NMTO downfolding of--
the spin-up LDA+U $a_{1g}$ band, we show that there are other hopping
integrals which are equally important for the band shape as the integral for
hopping between the partners of the pair.
\end{abstract}

\maketitle

A few years ago, Park et al. reexamined the socalled spin 1/2 model for
V$_{2}$O$_{3}$ \cite{Park00}. Based on polarization dependent x-ray absorption
measurements they showed that, for all phases, the vanadium $3+$ ion
$(d^{2})$ is in the spin 1 state. They also demonstrated that this state is
a mixture of $e_{g}^{\pi }e_{g}^{\pi }$ and $e_{g}^{\pi }a_{1g}$
configurations, with the former having the larger weight, especially at low
temperatures. Recall, that the $t_{2g}$ orbitals, which are $pd\pi$
anti-bonding with the O $p$ orbitals on the surrounding octahedron, lie
below the $pd\sigma $ anti-bonding $e_{g}$ orbitals and are split by a
trigonal distortion into low-lying, doubly degenerate $e_{g}^{\pi }$
orbitals and a higher-lying $a_{1g}$ orbital. The picture presented by Park
et al. is hardly consistent with the classical vertical-pair assumption that
the bonding-antibonding splitting of the $a_{1g}$ orbitals of the V-V pair
places the energy of the bonding orbital well \emph{below} that of the $%
e_{g}^{\pi }$ orbitals \cite{Mott70, McWhan73}. This is the assumption which
25 years ago led Castellani et al. to suggest the spin 1/2 model where for
the V-V pair two electrons fill the bonding $a_{1g}$ orbital and the two
remaining electrons 1/4-fill the four $e_{g}^{\pi }$ orbitals \cite%
{Castellani78}. It is now generally recognized that the spin 1/2 model is
incorrect. Nevertheless, the vertical pair remains a popular starting point
for current attempts to calculate the electronic structure of
V$_{2}$O$_{3}$ \cite{Mila00, Shina01, Joshi01}. A comprehensive review of
the latest experimental and theoretical results in this field can be found
in a recent paper by Di~Matteo, Perkins and Natoli \cite{Matteo02}.

\begin{figure}
\centering \includegraphics[clip=true,width=0.48\textwidth]{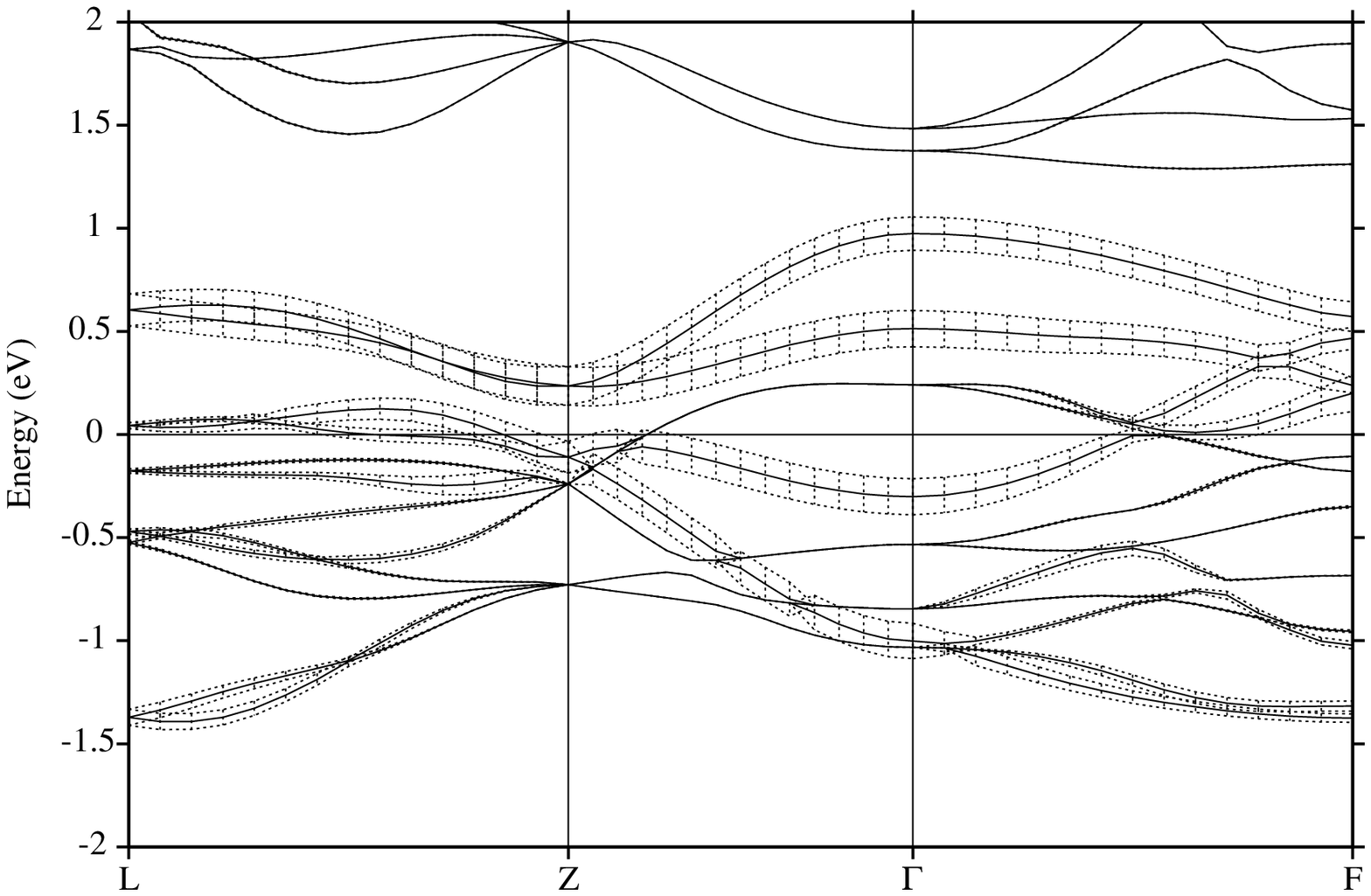}
\centering \includegraphics[clip=true,width=0.48\textwidth]{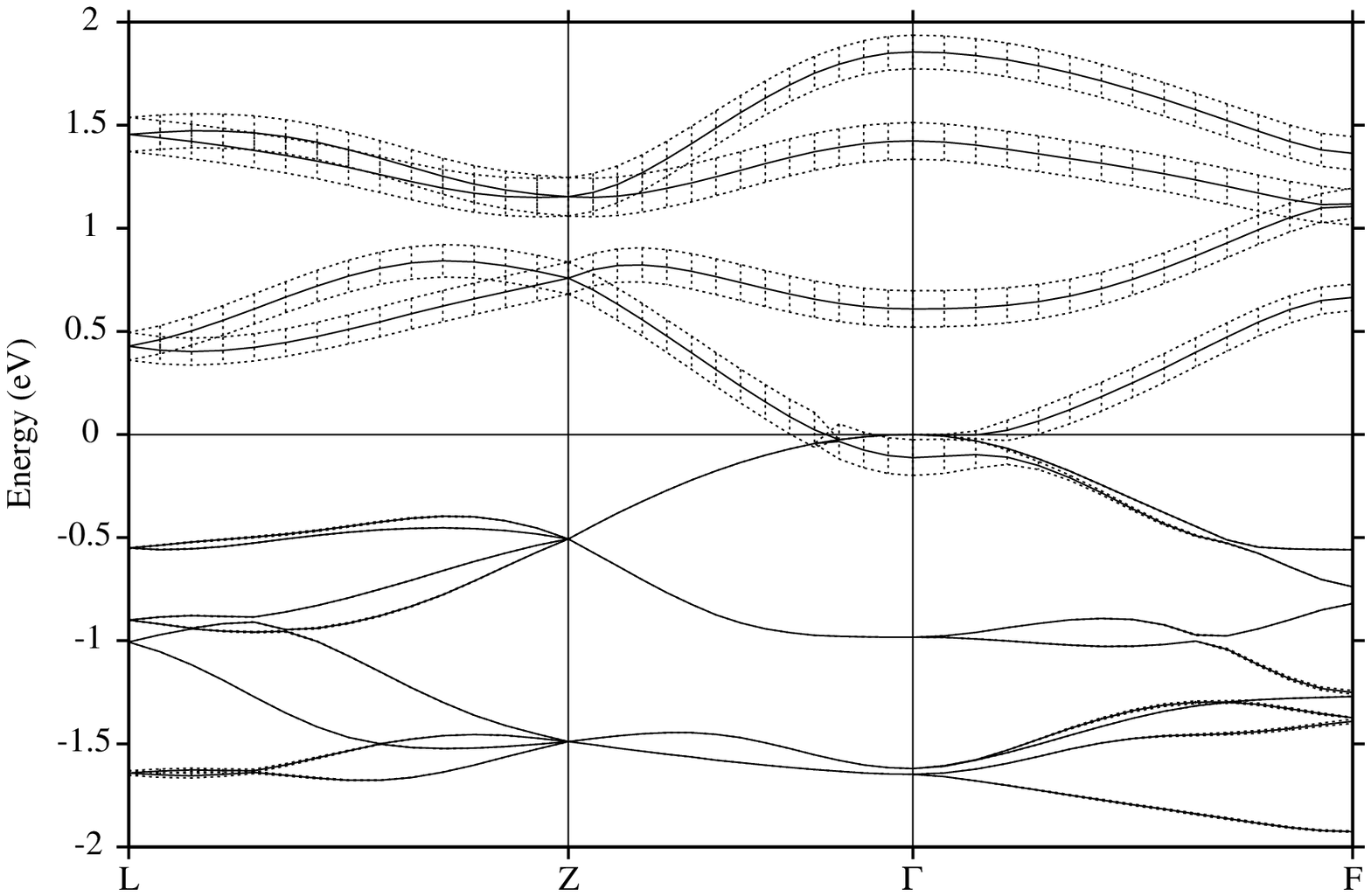}
\caption{LDA (upper panel) and LDA+U (lower panel) spin-up band structure of
ferromagnetic V$_{2}$O$_{3}$ in the corundum structure with 2 formula units
per cell. The amount of $a_{1g}$ character is indicated by
the width of so-called fat bands (dot-lines). We note that
$\Gamma$-Z is along the direction of the vertical pair (see Fig. 3.11 in
\cite{Bradley} and Fig.1 in \cite{Mattheiss94}). $\Gamma$=(0,0,0);
$Z$=(1/2,1/2,1/2); $L$=(0,1/2,0); $F$=(1/2,1/2,0).
The zero of energy is at Fermi energy.}
\label{fig1}
\end{figure}

In the present paper we study the dispersion of the $a_{1g}$ band obtained
from a modern LDA+U calculation by performing a tight-binding analysis.
Our motivation for doing this is to obtain information
regarding the smallest cluster that one can use in the model calculations while
still preserving the most important aspects of the band structure. In spite of
the fact that some literature exists providing the qualitative hint concerning
this, we considered it to be rather important to check this with more recent
methodology (e.g. the downfolding technique) providing the quantitative
estimates. Note that, although the spatial orientation of an $a_{1g}$ orbital
is actually in favour of a molecular-like picture, our analysis shows that
there are other hopping integrals which are equally important for description
of the main features of the band structure as the hopping integral between the
partners of the c axis pair (t$_1$). Though these inter-pair hopping integrals
are smaller than t$_1$, the contribution is proportional to the number of
neighbours and that makes them rather significant.

Since we are interested in understanding the relative importance of the
hopping between pairs and non-pairs, we consider the band structure of
ferromagnetic V$_{2}$O$_{3}$ in the high-temperature corundum structure.
Although this phase does not exist in nature, it can provide a good estimate
of an upper bound for the hopping integrals in this compound for the
following reasons: First of all, comparing with antiferromagnetic phases,
the ferromagnetic one has the largest band-width \cite{Ezhov99}. Secondly,
the distance between the partners of a vertical pair is shorter in the
corundum structure than in the low-temperature monoclinic phase
\cite{struc}. Therefore the hopping integral between partners of the pair
in this structure should be maximal.

\begin{figure}
\centering \includegraphics[clip=true,width=0.45\textwidth]{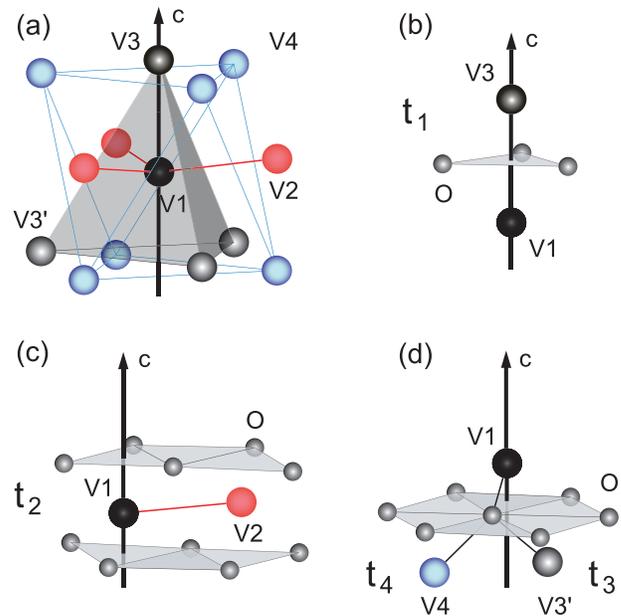}
\caption{Vanadium neighbors in corundum structure which are taken into
account in the tight-binding model (a). Definition of hopping parameters (b)
$t_{1}$ between the pair V1 and V3, (c) $t_{2}$ between V1 and V2, (d)
$t_{3}$ between V1 and V3$^\prime$ and $t_{4}$ between V1 and V4.
The small spheres denote the oxygen ions to illustrate the importance of
the bond angles.}
\label{fig2}
\end{figure}

In Fig.\ref{fig1} we show the LDA and LDA+U spin-up band structure of
V$_{2}$O$_{3}$ in the energy range of the 12 $t_{2g}$ bands
(4 V atoms per cell).
Comparing these two pictures one can see that the LDA+U band structure
calculated for U=3eV and J=0.8eV is essentially a rigid shift of the
$e_{g}^{\pi}$ band down in energy and $a_{1g}$ up so that the former is almost
completely below the Fermi energy and the later is above it. However, we note
that as a result the mixing between $e_{g}^{\pi}$ and $a_{1g}$ bands is
suppressed. Nevertheless, it is clear from Fig.\ref{fig1} that this mixing does
not come from the hybridization between different orbitals of the atoms in the
c-axis pair. This warns us already about the importance of the neighbours other
than the partner in the pair.

The LDA+U band structure, which yields a spin 1 $e_{g}^{\pi }e_{g}^{\pi}$ state,
has the advantage that, as already mentioned, the empty $a_{1g}$ band is
practically separated from the full $e_{g}^{\pi }$ band. Of course, this
depends on the values of the parameters U and J used in the calculations.
According to Solovyev et al. the calculated value of
the screened parameter U and Hund's rule exchange J for V 3+ ion in LaVO$_3$
are 3eV and 0.93eV, respectively \cite{Solovyev96}.
On the other hand, an empirical estimate by Marel and Sawatzky,
based on gas-phase multiplet splittings of the 3$d$ series, shows that
in the case of V 3$d$ J is about 0.74eV\cite{Marel88}.
In the present work we use J=0.8eV as estimated by Tanabe and Sugano
for the free V 3+ ion \cite{Tanabe54,Sugano}.

\begin{figure*}
\centering \includegraphics[clip=true,width=\textwidth]{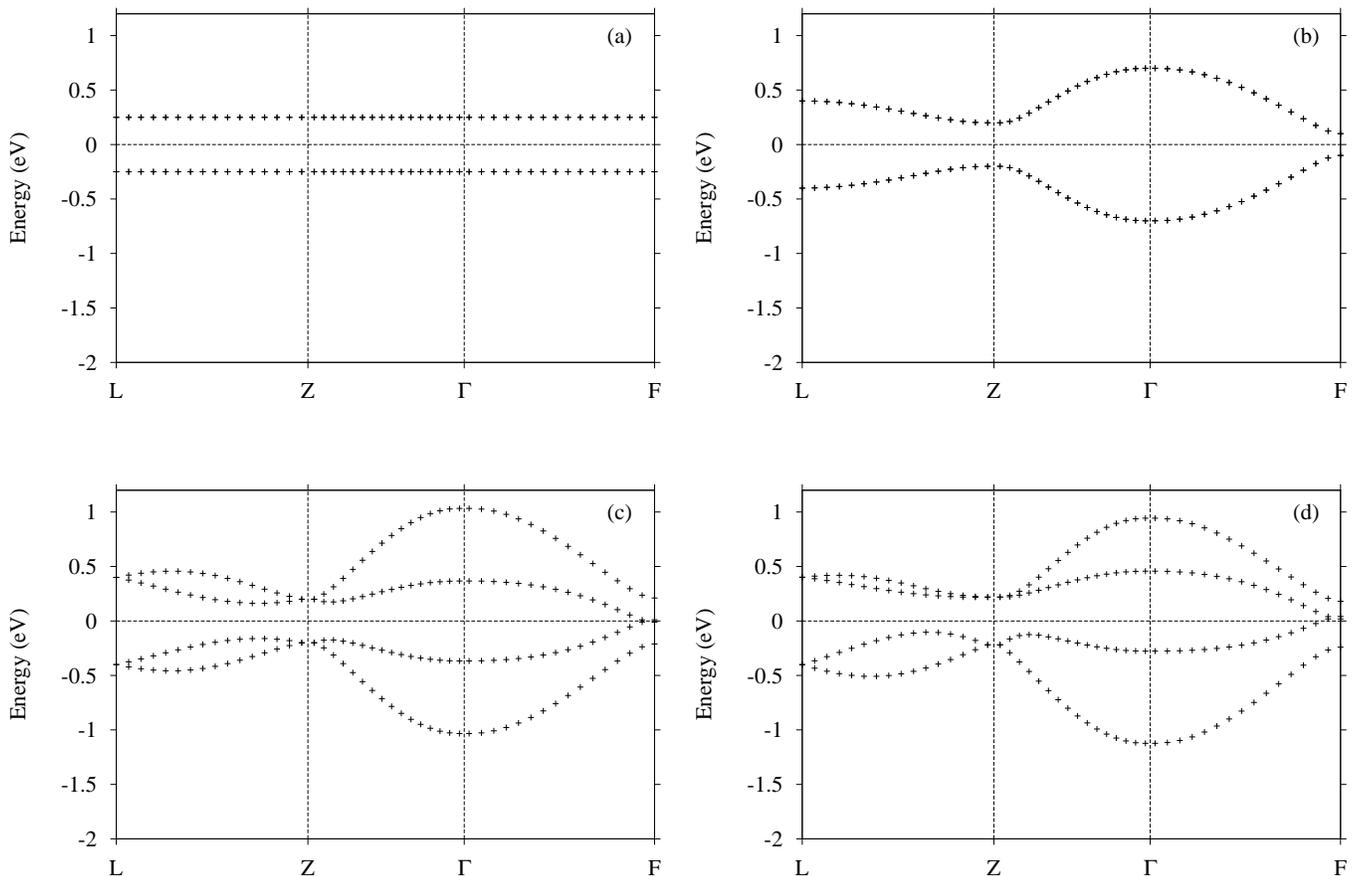}
\caption{Tight-binding $a_{1g}$ bands. The zero of energy is at the center
of the gap between the split bands. (a) $t_{1}$=--0.25eV, $t_{2}$=0, $t_{3}$%
=0, $t_{4}$=0. The splitting is $2\times t_{1}$ and there is no dispersion.
(b) $t_{1}$=--0.25eV, $t_{2}$=0, $t_{3}$=--0.15eV, $t_{4}$=0. Note that the
large splitting at $\Gamma $ is dominated by $t_{3}$ and not $t_{2}$. (c) $%
t_{1}$=--0.25eV, $t_{2}$=0, $t_{3}$=--0.15eV, $t_{4}$=--0.06eV. A small $%
t_{4},$ but with 6 nearest neighbors, is sufficient to cause the large
splitting into 4 bands seen in the LDA+U band structure. (d) $t_{1}$%
=--0.25eV, $t_{2}$=--0.03eV, $t_{3}$=--0.15eV, $t_{4}$=--0.06eV. To obtain
the very evident asymmetry between the upper and lower pairs of bands we
need a small $t_{2}$. This is now close to the LDA+U band structure. }
\end{figure*}

Before we discuss our results we note that, the width of the $a_{1g}$
band reaches its maximum at $\Gamma$, where it is about 2eV.
One might believe that this width is mainly caused by the bonding-antibonding
interaction between the vanadium pairs along the c-axis, and that the
interaction between pairs is weak. In the simplest such picture only two
hopping parameters would be important: A large intra-pair hopping parameter,
which should give the most of the bandwidth, and a smaller inter-pair
hopping. This would result in a situation where the splitting of the $a_{1g}$
band at the $\Gamma $-point is primarily determined by the value of
intra-pair hopping integral. However, one notices that the band splits into
\emph{four} almost \emph{equally} separated levels.

To shed more light on this issue, we carry out a tight-binding model
calculation where the hopping integrals to the first ($t_{1}$), second 
($t_{2}$), third ($t_{3}$) and fourth ($t_{4}$) nearest V neighbors are 
taken into account (Fig.\ref{fig2}).
Again, $t_{1}$ is the hopping integral between the atoms of the
c-axis pair. In Fig.3 (a) to (d) we demonstrate how each of these parameters
contribute in the dispersion of $a_{1g}$ band. Namely, switching on merely
the hopping parameter $t_{1}$ splits the atomic $a_{1g}$ level into two
doubly degenerate flat bands with energy difference $2t_{1}$. There is no
dispersion because the atoms in-between the pairs in the structure are
missing. In fact, only because these atoms are missing, does one see pairs
at all. Now the main dispersion is caused by the hopping parameter $t_{3}$
which yields a maximal splitting of $6t_{3}$ at the $\Gamma $-point (the
number of neighbors which an electron can hop to with $t_{3}$ is equal to
3). The hopping parameter $t_{4}$ lifts the degeneracy of each of these
doubly degenerate bands. Inclusion of $t_{2}$ makes the band asymmetric with
respect to the position of the initial $a_{1g}$ level. The final result in
Fig.3(d) looks very much like the LDA+U $a_{1g}$ band. The parameters used
are: $t_{1}$=--0.25eV, $t_{2}$=--0.03eV, $t_{3}$=--0.15eV, $t_{4}$=--0.06eV.
Although $t_{1}$ is indeed larger than other hopping integrals, its
influence on the bandwidth is not that large, because there is only one
nearest neighbour, as compared to 3 for $t_{2}$, 3 for $t_{3}$ and 6 for
$t_{4}$.

It is important to note that these values for the hopping parameters are not
unique. For example, $t_{1}$=--0.5eV, $t_{2}$=--0.03eV, $t_{3}$=--0.1eV and
$t_{4}$=--0.04eV would also give a small splitting at the Z-point, and large
one at $\Gamma $ or L, as is shown in Fig.4. This has to do with the
symmetry of these points. At the Z-point, for instance, the splitting
between the upper and lower components of the $a_{1g}$ band is $%
2|3t_{3}-t_{1}|,$ for $t_{2}$=0. Hence, for any given value of $t_{3}$ there
are always two values of $t_{1}\sim 3t_{3}\pm \delta _{Z}/2$ which give
exactly the same splitting. Note that $t_{4}$ does not influence the
energies at Z. At the L-point, on the other hand, the splitting is
determined primarily by the sum of $t_{1}$ and $t_{3}$. Therefore, \emph{no}
set of parameters with $t_{1}$ greater than 0.5eV can reproduce the LDA+U
band structure at this point (unless $t_{3}$ and $t_{1}$ have the opposite
signs, which is ''unphysical''). Comparing figures 1 and 4, one can easily
see that the two sets of parameters, which give the same splitting at Z,
will give different levels at F, and that only the set with $t_{1}=$--0.25
eV reproduces the accidental degeneray of the two middle levels in the
LDA+U. However, to reproduce the LDA+U band in such detail may not be
meaningful as long as all hopping integrals beyond $t_{4}$ are neglected.

\begin{figure}
\centering \includegraphics[clip=true,width=0.48\textwidth]{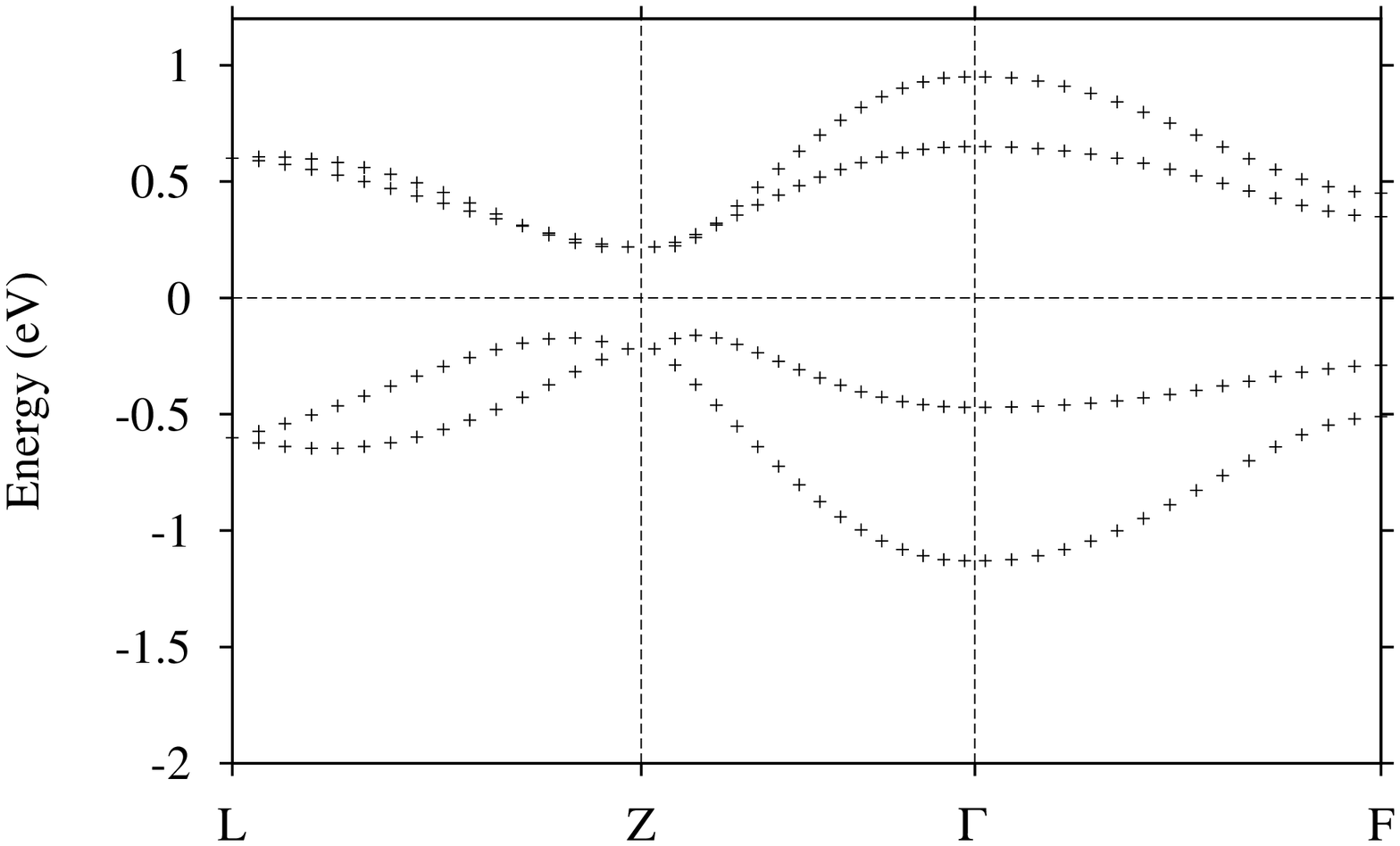}
\caption{Tight-binding $a_{1g}$ bands for $t_1$=--0.5eV, $t_2$=--0.03eV, 
$t_3$=--0.1eV and $t_4$=--0.04eV.}
\end{figure}

Although our tight-binding study has clearly demonstrated the role of the
various hopping integrals for the dispersion of the $a_{1g}$ band, and
although we can find parameters which fit the LDA+U $a_{1g}$ band, it is
difficult to select one set of hopping integrals because, to the accuracy
expected for our model, different sets can do this. The most
straightforward way to resolve this problem is to use the downfolding
procedure of Andersen \textit{et. al.} \cite{nmto}.
This procedure relies on keeping in
the first-principles NMTO band-structure calculation only the relevant
degrees of freedom, in this case the $a_{1g}$ Wannier-like orbitals whose
LDA+U spin-up bands lie in the energy range from the Fermi level to nearly 2
eV above, and integrating out the other degrees of freedom. This naturally
takes into account re-normalization effects due to the integrated-out
orbitals. Fourier transform of this few-orbital downfolded and symmetrically
orthonormalized NMTO Hamiltonian provides the hopping matrix elements of the
corresponding tight-binding Hamiltonian. This method provides a way of
generating Wannier-like functions and their single-particle Hamiltonian
without any fitting procedure. The detailed discussion of such calculations
for V$_{2}$O$_{3}$, as well as comparisons with Hamiltonians proposed
previously, will be presented elsewhere \cite{Tanusri}. Here we only mention
a technical point specific to the present application: Since the downfolding
procedure takes place at a more basic level than where U is ''added'' to the
LDA, we need to construct the \emph{potential} which yields the spin-up
LDA+U band structure. That potential we obtained from the LDA potential by
shifting its logarithmic-derivative functions, $\varphi _{Rlm}^{\prime
}\left( \varepsilon ,s\right) /\varphi _{Rlm}\left( \varepsilon ,s\right) ,$
in energy so as to reproduce the spin-up LDA+U band structure.

From this NMTO downfolding calculation we obtained the following hopping
integrals: $t_{1}$=--0.30eV, $t_{2}$=--0.02eV, $t_{3}$=--0.11eV and
$t_{4}$=--0.05eV, which are close, although not identical to those used in
Fig.3(d). As Fig.5(a) shows, the band structure obtained from these hopping
integrals differs a bit from that in Fig.3(d), and from the upper four LDA+U
bands in Fig.\ref{fig1}. The reason is simply that the downfolded band structure
shown in Fig.5(b), obtained by downfolding to the the four $a_{1g}$ bands,
cannot be reproduced completely with merely $t_{1},\,t_{2},\,t_{3},$ and
$t_{4}$. Its Hamiltonian has also non-zero higher Fourier components, which
is hardly surprising.
The downfolding calculation thus confirms the gross values of the hopping
integrals found by tight-binding fitting to the first 4 shells, but also
points to the need for including longer ranged hoppings to reproduce the
details.

\begin{figure}
\centering \includegraphics[clip=true,width=0.46\textwidth]{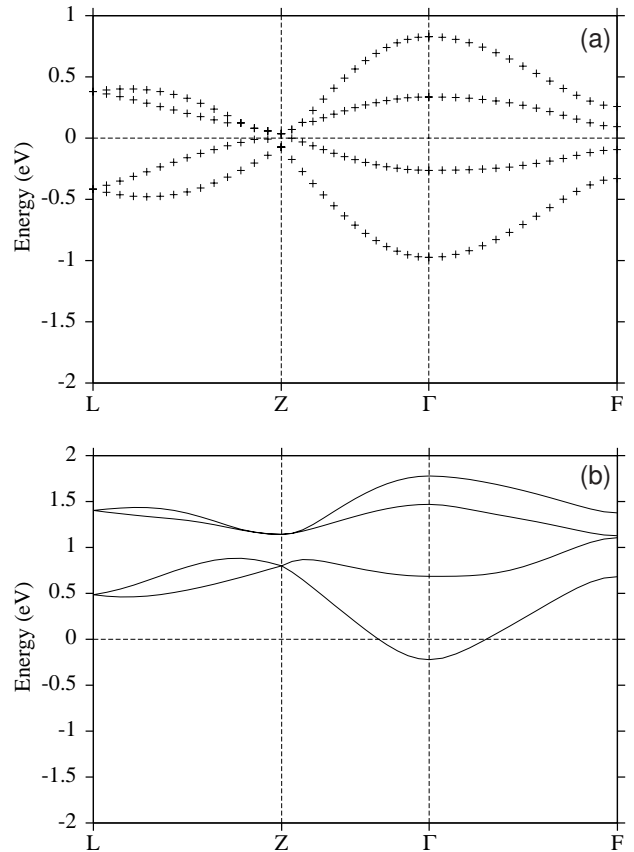}
\caption{Tight-binding $a_{1g}$ bands calculated with
$t_{1}$=--0.3eV, $t_{2}$=--0.02eV, $t_{3}$=--0.11eV and $t_{4}$=--0.05eV
(upper panel) and those obtained from the downfolding procedure (lower panel).
Note that the zero of energy in the upper panel corresponds to the energy
of atomic $a_{1g}$ level but it is at Fermi energy in the lower panel.}
\end{figure}

One should keep in mind that the hopping integrals discussed above are for
vanadium-centered Wannier-like orbitals and thus different from the Slater-
Koster hopping integrals for atomic oxygen and vanadium orbitals obtained by
Mattheiss \cite{Mattheiss94}. Following Harrison one can show that the hopping
integral between $a_{1g}$ atomic orbitals on the vanadium pair is about 0.8eV
\cite{Harrison}, whereas the one in which the oxygen degrees of freedom are
integrated out is much less than that. This reduction is due  to the
anti-bonding character of the $pd\pi$ interaction.

We have thus demonstrated that, although the integral for hopping between the
vertical pair is the largest hopping integral, it is not the single most
important one for the $a_{1g}$ bandwidth. This is so because the actual hopping
integrals are not only determined by the direct V-V hoping but also evolve via
intermediate O 2p orbitals. The simple picture where only the hopping parameter
within the c-axis pair is important is not sufficient to describe the $a_{1g}$
band in V$_{2}$O$_{3}$. Our calculations show that the hopping integrals
between second, third, and fourth nearest vanadium neighbors are equally
important.

The authors are grateful to O.K.~Andersen, V.I.~Anisimov and G.A.~Sawatzky
for very fruitful and inspiring discussions. I.S.E. is grateful for financial
support from the Spinoza prize program of the Netherlands Organisation for 
Scientific Research and from the Canadian Institute for Advanced research.
T.S.D. would like to  thank Kavli Institute of Theoretical Physics, Santa
Barbara for its hospitality during which a part of the work has been carried 
out. M.A.K. is grateful for financial support from the Russian Foundation for 
Basic Research RFFI-01-02-17063.

\end{document}